\documentclass[a4paper]{article}

\usepackage{INTERSPEECH2018}
\usepackage{subcaption}
\usepackage{color,xcolor}
\usepackage{multirow}
\usepackage{diagbox}
\usepackage{setspace}

\graphicspath{{figures/}}

\title{Investigating Generative Adversarial Networks based Speech Dereverberation for Robust Speech Recognition}

\name{Ke Wang$^{1,2}$, Junbo Zhang$^2$, Sining Sun$^1$, Yujun Wang$^2$, Fei Xiang$^2$, Lei Xie$^{1*}$\thanks{*Corresponding author}}
\address{
  $^1$Shaanxi Provincial Key Laboratory of Speech and Image Information Processing,\\
  School of Computer Science, Northwestern Polytechnical University, Xi'an, China\\
  $^2$Xiaomi, Beijing, China}
\email{\{kewang, snsun, lxie\}@nwpu-aslp.org, \{zhangjunbo, wangyujun, xiangfei\}@xiaomi.com}

\begin{document}
\begin{spacing}{0.80}

\maketitle
\begin{abstract}
We investigate the use of generative adversarial networks (GANs) in speech dereverberation for robust speech recognition. GANs have been recently studied for speech enhancement to remove additive noises, but there still lacks of a work to examine their ability in speech dereverberation and the advantages of using GANs have not been fully established. In this paper, we provide deep investigations in the use of GAN-based dereverberation front-end in ASR. First, we study the effectiveness of different dereverberation networks (the generator in GAN) and find that LSTM leads to a significant improvement as compared with feed-forward DNN and CNN in our dataset. Second, further adding residual connections in the deep LSTMs can boost the performance as well. Finally, we find that, for the success of GAN, it is important to update the generator and the discriminator using the same mini-batch data during training. Moreover, using reverberant spectrogram as a condition to discriminator, as suggested in previous studies, may degrade the performance. In summary, our GAN-based dereverberation front-end achieves 14\%$\sim$19\% relative CER reduction as compared to the baseline DNN dereverberation network when tested on a strong multi-condition training acoustic model.

\end{abstract}

\section{Introduction}\label{section:introduction}

The performance of automatic speech recognition (ASR) has been boosted tremendously in the last several years and state-of-the-art systems can even reach the performance of professional human transcribers in some conditions~\cite{xiong2017microsoft,kurata2017language}. However, room reverberation often seriously degrades the ASR performance, especially in far-field speech recognition where the talker is away from the microphone~\cite{kinoshita2016summary,wu2017reverberation}. Therefore, more attention has been paid recently in the research community to address this issue.

%MCT
%speech dereverberation .
%multi-channel signal processing
%Distance speech recognition can 
%in close-talk conditions

In theory, reverberant speech can be regarded as a room impulse response (RIR) convolving with the clean speech in the time domain~\cite{neely1979invertibility}. A straightforward approach is called \textit{speech dereverberation}, i.e., remove the reverberation from the contaminated speech. In this track, microphone array and multi-channel signal processing are very helpful~\cite{delcroix2007dereverberation,kumatani2012microphone}, but single-channel speech reverberation is still desirable in many real applications in which using multiple microphones may be impractical. Single-microphone speech dereverberation has been intensively studied in the signal processing community and a variety of approaches have been proposed~\cite{neely1979invertibility,wu2006two,kinoshita2009suppression,mosayyebpour2013single,mohammadiha2016speech}. Another approach to deal with reverberation (and noise) in speech recognition is multi-condition training (MCT), in which speech contaminated with reverberation, either simulated or real-recorded, is added in the acoustic model training set,  letting the model learn the reverberation effects automatically. Although the above approaches are reasonably effective, it is still far away from claiming success in the fight against reverberation in speech recognition.

%Based on this, a variety of one-microphone dereverberation techniques have been proposed by  traditional speech signal processing communities in the past. One idea to remove reverberant effects is by an inverse filter of RIR to deconvolve the reverberant speech~\cite{neely1979invertibility}. The work presented in~\cite{wu2006two} described a two-stage algorithm to increase signal-to-reverberant energy ratio and minimize the influence of long-term reverberation. Kinoshita \textit{et al.}~\cite{kinoshita2009suppression} suppressed the late reverberations using long-term multi-step linear prediction and spectral subtraction. In~\cite{mosayyebpour2013single}, an algorithm to blindly determine the inverse filter length and spectral subtraction was employed for dereverberation and denoising under very noisy reverberant conditions. Another approach is using non-negative matrix factorization (NMF)~\cite{lee2001algorithms} to remove reverberation~\cite{gemmeke2011exemplar,mohammadiha2016speech}. Nonetheless, almost all of these methods depend on some additional assumptions and can't be tightly integrated with acoustic models based on deep neural networks (DNNs).

Recently, due to their strong regression learning abilities, deep neutral networks (DNNs) have been used in speech enhancement~\cite{xu2014experimental} and later in speech dereverberation~\cite{kinoshita2016summary,wu2017reverberation,kinoshita2017derev}.  The deep structure can be naturally regarded as a dereverberation filter that can learn the essential relationship between the reverberant speech and its counterpart, i.e., the clean speech, through a set of multi-condition data. Various deep structures, e.g., feed-forward~\cite{han2015learning}, recurrent~\cite{weninger2014deep} and convolutional~\cite{park2016fully}, have been explored in the field. Either direct spectral mapping~\cite{wu2017reverberation,han2015learning} or masking~\cite{williamson2017time} can be considered in the dereverberation network. In the typical spectral mapping approach~\cite{xu2014experimental}, the multi-condition data set used in the network training usually consists of pairs of reverberant and clean speech represented by log-power spectra (LPS). Note that in speech recognition, the output of the dereverberation network can be features like FBanks or MFCCs, which do not need to be inverted back to waveforms.

%using log-power spectrum (LPS) feature and adopted a linear output layer instead of a nonlinear one~\cite{han2015learning}
%
%Recently, many studies preform dereverberation in a supervised manner based on DNNs due to their strong mapping capabilities. The work presented in~\cite{weninger2014deep} used deep bidirectional long short-term memory (LSTM) denoising auto-encoder (DAE) to do feature dereverberation. Han \textit{et al.}~\cite{han2015learning} proposed to dereverberate and denoise speech by learning a spectral mapping using feed-forward DNN. In~\cite{wu2017reverberation}, Wu \textit{et al.}\ showed a reverberation-time-aware (RTA) approach to speech dereverberation using log-power spectrum (LPS) feature and adopted a linear output layer instead of a nonlinear one~\cite{han2015learning}. Williamson \textit{et al.}~\cite{williamson2017time} performed dereverberation and denoising using time-frequency mask in complex domain. 

\begin{figure*}[t]
\centering
\begin{subfigure}[t]{0.3\textwidth}
  \centering
  \includegraphics[width=0.7\textwidth]{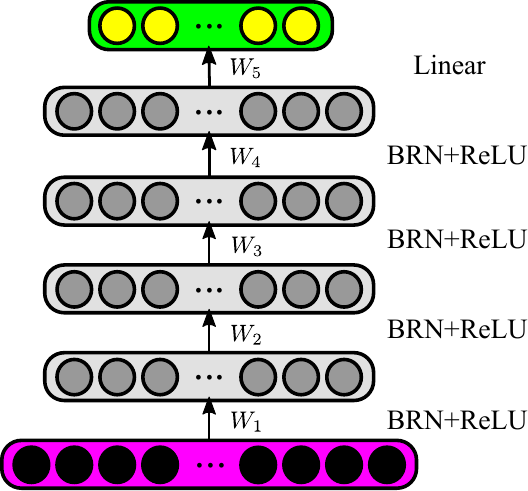}
  \subcaption{feed-forward DNN}
  \label{fig:dnn}
\end{subfigure}
\quad
\begin{subfigure}[t]{0.3\textwidth}
  \centering
  \includegraphics[width=1.0\textwidth]{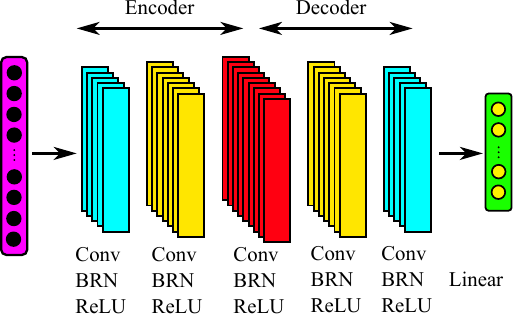}
  \subcaption{RCED}
  \label{fig:rced}
\end{subfigure}
\begin{subfigure}[t]{0.3\textwidth}
  \centering
  \includegraphics[width=0.4\textwidth]{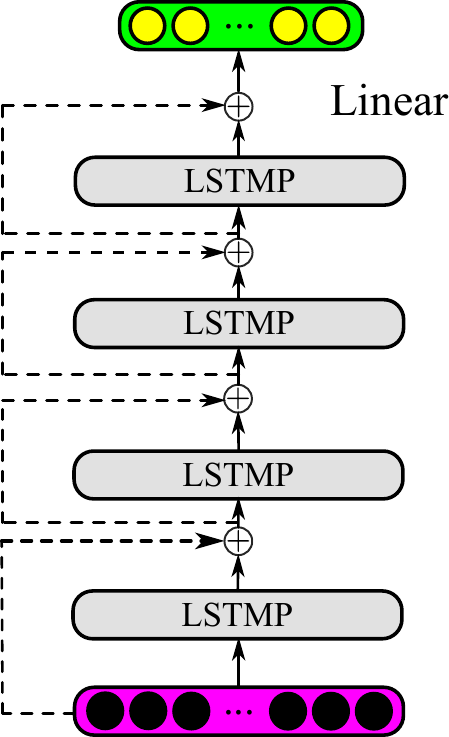}
  \subcaption{LSTM with layer-wise residue}
  \label{fig:lstm}
\end{subfigure}
%\caption{Architectures of the DNN based speech dereverberation models. (a) and (b) map spliced reverberant log-power spectra to clean MFCC features. The input of (c) is a single acoustic frame, and the dotted lines represent layer-wise residual, where the previous hidden activation is fed through residual connection.}
\caption{Architectures of different dereverberation networks.}
\label{fig:base_architecture}
\vspace{-0.4cm}
\end{figure*}

All the above DNN-based speech dereverberation approaches aim to minimize the mean square error (MSE) between the outputs of network and the ground truth. Hence, there is an underlying hypothesis that the enhanced speech has the minimal value in the MSE loss with the referenced clean speech. However, the MSE objective function has very strong implicit assumptions, e.g., independence of temporal or spatial, equal importance of all signal samples, and inaccurate to describe the degree of signal fidelity~\cite{wang2009mean}. To remedy this deficiency, generative adversarial networks (GANs)~\cite{goodfellow2014generative}, which consist of a generator network ($G$) and a discriminator network ($D$), learned through a min-max adversarial game, might be a good choice. Specifically, Pascual \textit{et al.} have recently demonstrated the promising performance of GAN in speech enhancement~\cite{pascual2017segan} in the presence of additive noise. In the SEGAN approach~\cite{pascual2017segan}, the generator $G$ tries to learn the distribution of the clean data and generate enhanced samples from noisy speech to fool the discriminator $D$; while $D$ aims to discriminate between the clean and enhanced samples (generated from $G$), which captures the essential difference between them.  While SEGAN works on the waveform level, which targets to improve the perceptual speech quality, Donahue \textit{et al.}~\cite{donahue2017exploring} have explored GAN-based speech enhancement for robust speech recognition. Specifically, in~\cite{donahue2017exploring}, GAN works on the log-Mel filter-bank spectra instead of waveforms. The results have shown that GAN enhancement improves the performance of a clean-trained ASR system on noisy speech but it falls short of the performance achieved by conventional MCT. By appending the GAN-enhanced features to the noisy inputs and retraining, a 7\% WER improvement relative to the MTR system was achieved.

%In the \textit{adversarial} game of GAN, $G$ tries to learn the distribution of the real data and generate samples to fool the discriminator; $D$ tries to capture the essential difference between the generated samples from $G$ and the real data. GANs have been recently explored in several regression tasks, e.g., speech synthesis~\cite{yang2017statistical} and speech enhancement~\cite{pascual2017segan}. Pascual \textit{et al.} have recently demonstrated promising performance of GAN in speech enhancement~\cite{pascual2017segan} in the presence of additive noise. The their SEGAN approach, GAN works directly on the waveform level (rather than feature level) yielded improvements to perceptual speech quality. However, SEGAN aimed to remove additive noise for communication rather than dereverberation for ASR tasks. 

While the major goal of the above GAN approaches is to remove additive noises, in this paper, we investigate the use of GANs in the mapping-based speech dereverberation for robust speech recognition. Although the same framework can be borrowed from these previous studies, we provide a series of deep investigations in the use of dereverberation front-end in ASR. First, we study the effectiveness of different dereverberation networks (used later as the GAN generator) and find that LSTM dereverberating network can achieve superior speech recognition performance as compared with feed-forward DNN and CNN. Second, further adding residual connections in the deep LSTMs can continuously boost the performance. Finally, we find that it is important to update the generator $G$ and the discriminator $D$ using the same mini-batch data during training for the success of GAN. Moreover, we discover that, using reverberant spectrogram as a condition to $D$, as suggested in~\cite{donahue2017exploring,michelsanti2017conditional}, may degrade the performance of $G$. In summary, using the dereverberation GAN can achieve 14\%$\sim$19\% relative character error rate (CER) reduction as compared with the DNN dereverberation baseline when tested on a strong multi-condition training acoustic model.

\section{Mapping based speech dereverberation}\label{section:mapping_methods}

Speech dereverberation can be achieved by a typical mapping approach~\cite{xu2014experimental}, in which a regression DNN (shown in Fig.~\ref{fig:dnn}) is trained by pairs of reverberant and clean LPS and a linear activation function at the output of DNN is adopted instead of a nonlinear one. Moreover, the target LPS feature is usually normalized globally over all training utterances into zero mean and unit variance (CMVN). In the dereverberation stage, the LPS features of input speech are fed into the well-trained regression DNN to generate the corresponding enhanced LPS features.  Finally, the dereverberated waveform is reconstructed from the predicted spectral magnitude and the reverberant speech phase with an overlap-add algorithm.

Besides LPS, the input and output of the dereverberation DNN can be other speech features, e.g., MFCC and FBank. The speech features do not need to be inverted back to waveforms, when used for robust ASR. In~\cite{han2015deep}, results show that the mapping from LPS to MFCC can achieve lower word error rate than the mapping from MFCC to MFCC in a speech recognition task under additive noise conditions. This also indicates that the transformation for different feature domains and nonlinear dereverberation function can be learned by the neural network simultaneously.  Furthermore, as shown in Fig.~\ref{fig:rced} and~\ref{fig:lstm}, CNN and LSTM can be used as enhancers as well. We expect that these more powerful network structures can bring further improvements in the speech dereverberation task. We will elaborate the network configurations and evaluate the performances of different networks later in Section~\ref{section:exp_mapping}.

%For the purpose of improving the performance, the environment information can be added to the system. Thus, RTA~\cite{wu2017reverberation} based algorithms was proposed to using two RT60-dependent parameters, namely frame shift size ($R$) and acoustic context size ($N$) separately. In the training stage, $R$ and $N$ depend on the utterance-level RT60, while an RT60 estimator and a lookup table are used to determine $R$ and $N$ in dereverberation stage.

\section{Dereverberation GAN}\label{section:gan}

\subsection{GAN}

\begin{figure}[tbp]
  \centering{}
  \includegraphics[width=0.6\linewidth]{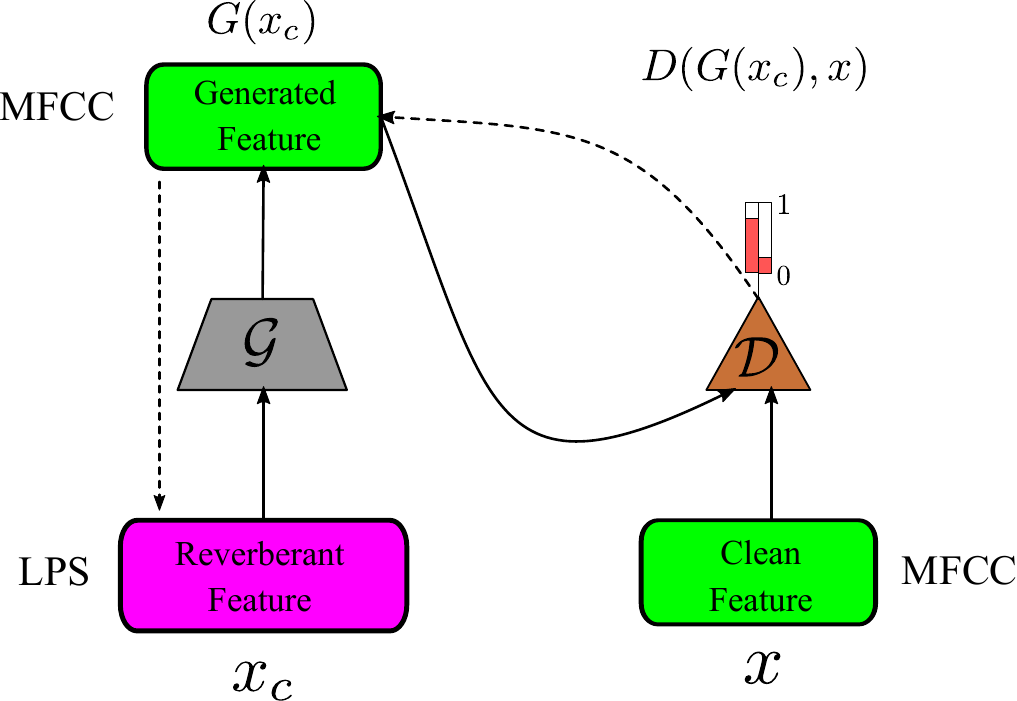}
  %\caption{System diagram of GAN based speech dereverberation framework. $G$ and $D$ can be any form of neural networks with or without residual connection. $D$ receives $G(x_c)$ or $x$ as input and decides whether the sample is real or dereverberated.}
  \caption{GAN based speech dereverberation framework.}
  \label{fig:gan}
  \vspace{-0.5cm}
\end{figure}

Generative adversarial networks (GANs)~\cite{goodfellow2014generative} are generative models implemented by two neural networks competing with each other in a two-player min-max game. Specifically, the generator network $G$ tries to learn a distribution $P_{g}(x)$ over data $x$ and a prior input noise variables $p_z(z)$. The aim is to match the true data distribution $P_{data}(x)$ to fool the discriminator $D$. The discriminator network $D$ serves as a binary classifier which aims to determine the probability that a given sample comes from the real dataset rather than $G$. Because of the weak guidance, the vanilla generative model cannot generate desirable samples. Hence the conditional GAN (CGAN)~\cite{mirza2014conditional} was proposed to steer the generation by considering extra information $x_c$ with the following objective function:
\begin{footnotesize}
\begin{equation}\label{eq:cgan}
\begin{split}
  \min_G\max_D V(G,D) = \mathbb{E}_{x\sim p_{data}(x,x_c)}[logD(x,x_c)] \\[-0.1cm]
                      + \mathbb{E}_{x_c\sim p_{data}(x_c),z\sim p_z(z)}[log(1-D(G(z,x_c),x_c))]\,.
\end{split}
\vspace{-0.5cm}
\end{equation}
\end{footnotesize}

In order to stabilize training and increase the quality of the generated samples in $G$, least-squares GAN (LSGAN)~\cite{mao2016least} was further proposed and the objective function changes to \vspace{-0.15cm}
\begin{footnotesize}
\begin{equation}\label{eq:lsgan_d}
\begin{split}
  %\min_{D}V_{\text{LSGAN}}(D) = \frac{1}{2}\mathbb{E}_{x\sim p_{data}(x,x_c)}[(D(x,x_c)-1)^2] \\[-0.2cm]
  %                            + \frac{1}{2}\mathbb{E}_{x_c\sim p_{data}(x_c),z\sim p_z(z)}[D(G(z, x_c))^2]\,,
  \min_{D}V(D) = \frac{1}{2}\mathbb{E}_{x\sim p_{data}(x,x_c)}[(D(x,x_c)-1)^2] \\[-0.2cm]
                              + \frac{1}{2}\mathbb{E}_{x_c\sim p_{data}(x_c),z\sim p_z(z)}[D(G(z, x_c), x_c)^2]\,,
\end{split}
\end{equation}
\vspace{-0.15cm}
\end{footnotesize}
\begin{footnotesize}
\begin{equation}\label{eq:lsgan_g}
  %\min_{G}V_{\text{LSGAN}}(G) = \frac{1}{2}\mathbb{E}_{x_c\sim p_{data}(x_c),z\sim p_z(z)}[(D(G(z,x_c))-1)^2]\,.
  \min_{G}V(G) = \frac{1}{2}\mathbb{E}_{x_c\sim p_{data}(x_c),z\sim p_z(z)}[(D(G(z,x_c), x_c)-1)^2]\,.
\end{equation}
\vspace{-0.3cm}
\end{footnotesize}

\subsection{Speech dereverberation with GAN}
It is straightforward to use GAN in speech dereverberation and Fig.~\ref{fig:gan} illustrates such a kind of architecture. It consists of a $G$ and a $D$, where $G$, serving as the mapper in conventional methods, tries to learn a transformation from reverberant speech to clean speech and $D$ tries to determine whether the input samples come from $G(x_c)$  or real-data $x$. Similar to~\cite{han2015deep}, $G$ aims to learning a mapping from the LPS feature input to the MFCC feature output which can be directly used in ASR. In some works~\cite{donahue2017exploring,michelsanti2017conditional}, the latent code $z$ is excluded from the generator $G$ to learn a direct mapping instead of a diversified translation in the original image-to-image translation task~\cite{isola2016image}. We borrowed this idea, but we remove the reverberant spectrogram as a condition to $D$. As we will report in Section~\ref{section:exp_gan}, the added reverberant spectrogram as a condition to $D$ not only increases the parameter size of $D$, but also degrades the performance of $G$. Therefore, we learn a generator distribution $P_g(x)$ over the conditional data $P_{data}(x_c)$ with the following proposed objective function:\vspace{-0.2cm}
\begin{footnotesize}
\begin{equation}\label{eq:lsgan_d_new}
\begin{split}
  \min_{D}V(D) &= \frac{1}{2}\mathbb{E}_{x\sim p_{data}(x)}[(D(x)-1)^2] \\[-0.1cm]
               &+ \frac{1}{2}\mathbb{E}_{x_c\sim p_{data}(x_c)}[D(G(x_c))^2]\,,
\end{split}
\end{equation}
\vspace{-0.2cm}
\end{footnotesize}
\begin{footnotesize}
\begin{equation}\label{eq:lsgan_g_new}
  \min_{G}V(G) = \frac{1}{2}\mathbb{E}_{x_c\sim p_{data}(x_c)}[(D(G(x_c))-1)^2]\,.
\end{equation}
\vspace{-0.15cm}
\end{footnotesize}

%which consists of a $G$ and a $D$. $G$ tries to learn a transformation from reverberant spectrogram to anechoic MFCC feature without Gaussian noise $z$. $D$ tries to determine whether the inputs sample comes from $G(x_c)$  or $x$ without any conditions. Then the adversarial loss (Equation~\ref{eq:lsgan_g_new}) and the MSE loss between $G(x_c)$ and $x$ are added with different weights to optimize the parameters of $G$. In dereverberated stage, we can easily generate anechoic feature from $G$. 

%Isola \textit{et al.}~\cite{isola2016image} reported that adding noise to $G$  was not effective.  Then they provided noise in the form of dropout but failed to produce stochastic output. However,  we are more concerned about accurate transformation between reverberant spectrogram and clean one in speech dereverberation tasks than learning a diversified translation in image-to-image tasks. Meanwhile, as we know, reverberant spectrogram is a high dimensional matrix and contains much information about clean signal. Inspired by these, we don't add $z$ but only use reverberant spectrogram as an input to $G$ which is similar with~\cite{michelsanti2017conditional}.

To further improve the ability of the adversarial component, previous CGAN approaches have indicated that it is beneficial to mix the GAN objective function with some numerical loss functions~\cite{mirza2014conditional}. We follow this approach in the dereverberation GAN approach and the MSE loss is controlled by a new hyper-parameter $\lambda$. Finally Eq.~(\ref{eq:lsgan_g_new}) becomes\vspace{-0.15cm}
\begin{footnotesize}
\begin{equation}\label{eq:lsgan_mse}
\begin{split}
  \min_{G}V(G) &= \frac{1}{2}\mathbb{E}_{x_c\sim p_{data}(x_c)}[(D(G(x_c))-1)^2] \\[-0.1cm]
                              &+ \frac{1}{2} \lambda \mathcal{L}_{\text{MSE}}(G(x_c), x)\,.
\end{split}
\end{equation}
\vspace{-0.15cm}
\end{footnotesize}

In practice, the generator $G$ can be a feed-forward network, a convolutional network or a LSTM RNN network, as we described in Section~\ref{section:mapping_methods}. Note that the discriminator $D$ is only used in the training and discarded in the dereverberation stage. In our approach, a 2-layer LSTM without residual connection is set to be the architecture of $D$.

\section{Experiments and results}\label{section:experiment}

\begin{table}[tp]
  \footnotesize
  \caption{CERs (\%) of Clean and MCT acoustic models.}
  \label{tab:am}
  \centering
  \scalebox{1.0}[1.0]{
    \begin{tabular}{*{6}{c}}
      \toprule
      \diagbox{AM}{Test} & \textbf{Clean} & \textbf{Real}  & \textbf{Simu} \\
      \hline
      {Clean}   & $7.86$ & $23.85$ & $20.24$ \\
      {MCT}     & $7.81$ & $16.02$ & $13.99$ \\
      \bottomrule
    \end{tabular}
  }
  \vspace{-0.6cm}
\end{table}

\subsection{Datasets}
In the experiments, we used a Mandarin corpus as our source of clean speech data, which consists of 103,000 utterances (about 100hrs). The RIRs were from~\cite{ko2017study}, including real-recorded RIRs and simulated RIRs for small, medium and large rooms. We randomly selected 97,000 utterances for network training and 3000 utterances for validation, and convolved with the RIRs (both real-recorded and simulated) to obtain the reverberant utterances. The rest 3000 utterances were used for testing and convolved with the real RIR and the simulated RIRs for small, medium and large rooms. Finally we obtained a testing set named `Real' that contains 3000 reverberant speech utterances convolved with real RIRs and another testing set named `Simu' that contains 9000 reverberant speech utterances convolved with simulated RIRs (3000 for  small/medium/large). To test the generalization ability of our approach, we ensured the RIRs used for training and testing were totally different.  All waveforms were sampled at 16\,kHz. We used Kaldi~\cite{povey2011kaldi} to generate the reverberant speech by convolving the clean signal with the corresponding RIR. As for feature extraction, the frame length was set to 25\,ms with a frame shift of 10\,ms.

\subsection{ASR back-end}
Our speech dereverberation front-end was used for speech recognition experiments. We used Kaldi to train our back-end ASR system with the similar acoustic model architecture and features in~\cite{peddinti2015time}.  The original training dataset consists of 1600\,hrs Mandarin speech data. We used speed-perturbation and volume-perturbation techniques~\cite{ko2015audio} to do data augmentation. Hence the clean model were trained using 4800\,hrs of speech data (1600~$\times$~3). We also trained an acoustic model (AM) using multi-condition training (MCT) strategy. The training data for the MCT model is 6400\,hrs (1600~$\times$~4), including the above 4800\,hrs of clean data and 1600\,hrs of reverberant data generated by convolving the clean data with the RIRs in~\cite{ko2017study} as the dereverberation front-end.

The time delay neural network (TDNN) acoustic model (AM) had 6 layers, and each layer had 850 rectified linear units (ReLUs) with batch renormalization (BRN)~\cite{ioffe2017batch}. The input contexts of TDNN AM were set to [$-$2,2]-\{$-$1,2\}-\{$-$3,3\}-\{$-$7,2\}-\{$-$3,3\}-\{0\} and the output softmax layer had 5795 units. The notation [$-$2,2] means we splice together frames $t-2$ through $t+2$ at the input layer and the notation \{$-$1,2\} means we splice together the input at the current frame minus 1 and the current frame plus 2. The input of the AM was 40-dimensional MFCC. All the speech dereverberation front-ends were tested on both Clean and MCR AMs. A trigram language model (LM), which was trained on about 2\,TB scripts with more than 100,000 words in the vocabulary, was used for decoding in the experiments. We also used entropy-based parameter pruning~\cite{stolcke2000entropy} and the threshold was set to be $10^{-8}$. 

The baseline results of Clean and MCT model are shown in Table~\ref{tab:am}. We can see a significant increase in CER when speech is contaminated with reverberations. In extending the training data of acoustic model by adding reverberant speech, the MCT AM can greatly reduce CER. 

\subsection{Mapping-based speech dereverberation}\label{section:exp_mapping}

We first investigated the speech dereverberation performances of different networks and input features in the mapping-based approach. Later we will select the best network as the generator in the GAN-based speech dereverberation. Specifically, we tested three different dereverberation networks, i.e., feed-forward DNN, redundant convolutional encoder decoder (RCED) and LSTM. As shown in Fig.~\ref{fig:base_architecture}, the DNN has 4 hidden layers and each of which contains 1024 ReLU neurons. The structure of the RCED is similar with~\cite{park2016fully} except the last layer. We changed the last filter CNN layer to a fully connected output layer as shown in Fig.~\ref{fig:rced}, because our input and target features were not in the same dimension. The input feature contains a context window of 11 frames ($t\pm 5$) for the DNN and the RCED. The number of filters and filter width of RCED model were set to 12-16-20-24-32-24-20-16-12 and 13-11-9-7-7-7-9-11-13, respectively. The learning rate was set to 0.001 with a mini-batch size of 256. Moreover, BRN was also used for DNN and RCED training. Instead of using vanilla LSTM, we adopted an LSTM with recurrent projection layer (LSTMP)~\cite{sak2014long}, which means we do not need to add an extra layer to do residual add like sDNN2 in~\cite{tu2017speech} to avoid dimension mismatch. The LSTM has 4 LSTMP layers followed by a linear output layer. Each LSTMP layer has 760 memory cells and 257 projection units and the input to the LSTM is a single acoustic frame. The learning rate was set to 0.0003 and the model was trained with 8 full-length utterances parallel processing. 

%In our front-end dereverberation experiments, we use 257-dimensional LPS, which is consistent with~\cite{wu2017reverberation,xu2014experimental,han2015deep}, as input feature and 40-dimensional MFCC as target feature. As shown in Table~\ref{tab:baseline}, LPS is robuster than MFCC, which is consistent with~\cite{han2015deep}. In addition, CMVN is applied to the input and target feature.

\begin{table}[tbp]
  \footnotesize
  \caption{CERs (\%) of different front-end networks.}
  \label{tab:baseline}
  \centering
  \scalebox{1.0}[1.0]{
    \begin{tabular}{*{7}{c}}
      \toprule
      \multirow{2}{*}{\textbf{Input}} & \multirow{2}{*}{\textbf{Method}} &  \multicolumn{2}{c}{\textbf{Clean AM}}  & \multicolumn{2}{c}{\textbf{MCT AM}} \\
                                      &                                    & \textbf{Real} & \textbf{Simu}           & \textbf{Real} & \textbf{Simu} \\
      \midrule      
      \multirow{3}{*}{MFCC} & DNN  & $17.86$ & $16.63$ & $16.31$ & $14.72$ \\
                            & RCED & $18.28$ & $16.73$ & $16.76$ & $15.09$ \\
                            & LSTM & $15.38$ & $13.37$ & $14.21$ & $12.46$ \\
      \hline
      \multirow{3}{*}{LPS}  & DNN  & $16.62$ & $15.33$ & $15.35$ & $14.03$ \\
                            & RCED & $15.55$ & $14.15$ & $14.15$ & $13.09$ \\
                            & LSTM & $15.04$ & $13.16$ & $13.97$ & $12.20$ \\
      \bottomrule
    \end{tabular}
  }
\end{table}

\begin{table}[tbp]
  %\caption{CERs (\%) comparisons for different layers and residual connection architectures. ResNet-I and ResNet-L stand for input residue connection and layer-wise residue connection respectively.}
  \footnotesize
  \caption{CER (\%) comparisons for different layers and residual connection architectures.}
  \label{tab:resnet}
  \centering
  \scalebox{1.0}[1.0]{
    \begin{tabular}{lcccccc}
      \toprule
      \multirow{2}{*}{\textbf{Method}} &  \multicolumn{2}{c}{\textbf{Clean AM}}  & \multicolumn{2}{c}{\textbf{MCT AM}} \\
                                       &  \textbf{Real} & \textbf{Simu}          & \textbf{Real} & \textbf{Simu} \\
      \midrule
      2-layer LSTM           & $\bm{15.41}$ & $\bm{13.50}$ & $\bm{14.25}$ & $\bm{12.55}$ \\
      \ \textit{+ Res-I} & $16.18$ & $14.41$ & $14.99$ & $13.06$ \\
      \ \textit{+ Res-L} & $16.13$ & $13.74$ & $14.61$ & $12.65$ \\
      \hline
      4-layer LSTM           & $15.04$ & $13.16$ & $13.97$ & $\bm{12.20}$ \\
      \ \textit{+ Res-I} & $15.81$ & $13.48$ & $14.60$ & $12.47$ \\
      \ \textit{+ Res-L} & $\bm{14.99}$ & $\bm{13.13}$ & $\bm{13.90}$ & $12.22$ \\
      \hline
      8-layer LSTM           & \multicolumn{4}{c}{divergence}\\
      \ \textit{+ Res-I} & $15.53$ & $13.55$ & $14.48$ & $12.49$ \\
      \ \textit{+ Res-L} & $\bm{14.67}$ & $\bm{12.75}$ & $\bm{13.62}$ & $\bm{12.04}$ \\
      \bottomrule
    \end{tabular}
  }
  \vspace{-0.4cm}
\end{table}

\begin{table*}[tbp]
  %\caption{CERs (\%) comparisons by previous mapping based networks and our proposed framework. ``DB'' means we use different mini-batch data to update the parameters of GAN and ``CD'' means we add the conditional information to input of $D$. Relative improvements are given in parentheses w.r.t. the corresponding DNN model.}
  \footnotesize
  \caption{CERs (\%) comparisons by previous mapping based networks and our proposed framework. ``DB'' means we use different mini-batch data to update the parameters of GAN and ``CD'' means we add the conditional information to input of $D$. Relative improvements are given in parentheses w.r.t. the corresponding DNN model.}
  \label{tab:res_gan}
  \centering
  \scalebox{1.0}[1.0]{
    \begin{tabular}{lccccc}
      \toprule
      \multirow{2}{*}{\textbf{Method}} &  \multicolumn{2}{c}{\textbf{Clean AM}}  & \multicolumn{2}{c}{\textbf{MCT AM}} \\
                                       &  \textbf{Real} & \textbf{Simu}          & \textbf{Real} & \textbf{Simu} \\
      \midrule
      SEGAN                    & $32.98~(-98.44)$ & $37.14~(-142.27)$ & $30.18~(-96.61)$ & $32.37~(-130.72)$ \\
      \hline
      DNN                      & $16.62~(0.00)$   & $15.33~(0.00)$    & $15.35~(0.00)$   & $14.03~(0.00)$    \\
      %\ \emph{+ RTA}           & $16.66~(-0.24)$  & $15.25~(0.52)$    & $15.46~(-0.72)$  & $14.10~(-0.50)$ \\
      %\hline
      LSTM                     & $15.04~(9.51)$   & $13.16~(14.16)$   & $13.97~(8.99)$   & $12.20~(13.04)$ \\
      %\ \emph{+ RTA}           & $14.92~(10.23)$  & $13.14~(14.29)$   & $13.86~(9.71)$   & $12.29~(12.40)$ \\
      \ \textit{+ Res}           & $14.99~(9.81)$   & $13.13~(14.35)$   & $13.90~(9.45)$   & $12.22~(12.90)$ \\
      \ \textit{+ GAN}           & $\bm{14.07~(15.34)}$ & $12.02~(21.59)$ & $13.15~(14.33)$ & $11.42~(18.60)$ \\
      % \hline
      % LSTM+GAN+Res                  & $14.10~(15.16)$  & $\bm{11.96~(21.98)}$ & $\bm{13.14~(14.40)}$ & $\bm{11.40~(18.75)}$ \\
      %\ \emph{+ RN}            & $14.71~(11.49)$  & $12.66~(17.42)$   & $13.77~(10.29)$  & $11.86~(15.47)$ \\
      \ \textit{+ GAN+Res}              & $14.10~(15.16)$  & $\bm{11.96~(21.98)}$ & $\bm{13.14~(14.40)}$ & $\bm{11.40~(18.75)}$ \\
      \ \textit{+ GAN+Res (DB)}          & $15.72~(5.42)$   & $13.95~(9.00)$   & $14.60~(4.89)$    & $12.83~(8.55)$ \\
      \ \textit{+ GAN+Res+CD}          & $14.27~(14.14)$  & $12.19~(20.48)$   & $13.38~(12.83)$  & $11.43~(18.53)$ \\
      \bottomrule
    \end{tabular}
  }
  \vspace{-0.5cm}
\end{table*}

All the models explored here were optimized with the Adam~\cite{kingma2014adam} method and initialized with the Xavier~\cite{glorot2010understanding} algorithm. We also used exponential decay to decrease the learning rate which was similar with Kaldi nnet3\footnote{egs/wsj/s5/steps/libs/nnet3/train/common.py(get\_learning\_rate)} and the terminated learning rate was 5 orders of magnitude smaller than the initial learning rate.

In Table~\ref{tab:baseline}, we list all experimental results on both Clean and MCT AMs. Firstly, we observe consistent improvement on all dereverberation networks by replacing MFCC with LPS features as the network input. Here the LPS feature is 257 dimension and the MFCC feature is 40 dimensions. Note that the output of all the dereverberation networks is 40-dimension MFCC which is fed into the ASR system. This conclusion is consistent with that in~\cite{han2015deep}, where LPS performs better than MFCC when used as the input of a denoising network.

When we compare Table~\ref{tab:baseline} with Table~\ref{tab:am}, we can find that the mapping-based dereverberation works quite well. When tested on the Clean AM, all the dereverberation networks are effective with significant CER reduction; when tested on the MCT AM, the dereverberation networks with the LPS input are still effective with apparent CER reduction. Comparing different model structures, we discover that LSTM achieves the best performance. For instance, the LPS-LSTM dereverberation network reduces the CER from 23.85\% (real-reverberation added) to 15.04\% for the Clean AM and reduces the CER from 16.02\% (real-reverberation added) to 13.97\% for the MCT AM. We believe that the superior peformance is because of the LSTM's ability to model long-term contextual information that is essential is the speech dereverberation task. We also find RCED-CNN is not good when MFCC is used as the input. We will use LSTM as our network in the rest of the experiments.

%at in speech dereverberation task for low-dimensional feature by comparing the results of MFCC and LPS as input feature. Thus, our following GAN and ResNet based experiments are all based on LSTM.
%In order to compare our works with previous excellent researches, we also use RTA to further improve the performance of front-end enhancer. Unlike vanilla RTA methods, we splice RT60 to input feature because LSTM don't need contextual information and we don't convert enhanced feature to waveform and our enhanced features are evaluated on fixed AMs. Otherwise, we need train different AMs with different frame shift size ($R$) and acoustic context size ($N$). Results are shown in Table~\ref{tab:res_gan}. We can find that our results of DNN-RTA is worse than DNN but LSTM-RTA is better than LSTM. That means $R$'s and $N$'s selections are important for DNN and RT60 information is important to LSTM to convey the conveyance of hidden state.

\subsection{Adding ResNet}

Table~\ref{tab:resnet} shows the results of different residual connection architectures. The layer-wise residual connection (Res-L) structure can be seen in Fig.~\ref{fig:lstm}; while the input residual connection (Res-I) structure is similar with Res-L and more details can be found in~\cite{chen2017improving}.  As we expected, it's not necessary to add residual connections to shallow networks. Performances degrade when residual connections are used in a 2-layer LSTM. Res-L always performs better than Res-I. This is reasonable because Res-L tries to learn the residue of the high-level abstract feature while Res-I just learns the residue of the input feature. When the LSTM is as deep as 4 layers, Res-L starts to work and the lowest CERs are achieved when the LSTM has 8 layers. As training a 8-layer LSTM is time-consuming, we perform the GAN experiments with a 4-layer LSTM generator in the following.

%When the number of hidden layers reaches to more than 4, ``ResNet-L'' is always better than ``ResNet-I'', because ``ResNet-L'' tries to learn the residue of high-level abstract feature but ``ResNet-I'' just learns the residue of input feature. It can also be found that ResNet is an important element for mapping based speech dereverberation methods, especially for deep architectures.

\subsection{Speech dereverberation with GAN}\label{section:exp_gan}

We finally investigated the ability of GAN in mapping-based speech dereverberation. We also reproduced the SEGAN approach~\cite{pascual2017segan} with the open-source codes\footnote{https://github.com/santi-pdp/segan} as a comparison. As shown in Table~\ref{tab:res_gan}, SEGAN degrades the ASR performance within our expectation, which is consistent with the reported results in~\cite{donahue2017exploring}. We believe this is because SEGAN aims to improve the perception of noisy speech and time-domain enhancement may be not appropriate for reverberant speech recognition. 

%In order to make comparisons, we also reimplemented the recent RTA  approach [x]. We also introduced the RTA to an LSTM network. We spliced RT60 to the input feature because LSTM does not need contextual information and we don't convert enhanced feature to waveform and our enhanced features are evaluated on fixed AMs.  [???] Otherwise, we need to train different AMs with different frame shift size ($R$) and acoustic context size ($N$). Results are shown in Table~\ref{tab:res_gan}. We can find that our results of DNN-RTA are slightly worse than DNN but LSTM-RTA is better than LSTM. This means the selection of $R$'s and $N$'s is important for DNN and the RT60 information is essential to LSTM to convey the conveyance [???] of the hidden state.

In the proposed GAN-based methods, the architecture of $G$ is consistent with that in Fig.~\ref{fig:lstm} with 4 hidden LSTMP layers. The architecture of $D$ is similar with $G$ but contains only 2 LSTMP layers and the cell number and the projection dimension are set to 256 and 40, respectively. The hyper-parameter $\lambda$ in Eq.~(\ref{eq:lsgan_mse}) was set to 200 and the learning rate of $G$ and $D$ were set to $0.00008$ and $0.0003$, respectively. In each iteration, we updated the parameters of $G$ twice and the parameters of $D$ once.  To stabilizing GAN training, we also add instance Gaussian noise to the MFCC input of $D$\footnote{http://www.inference.vc/instance-noise-a-trick-for-stabilising-gan-training/}. In Table~\ref{tab:res_gan}, we demonstrate that using GAN (in LSTM+GAN) is not only viable but also outperforms the LSTMs. 

At the early stage of our experiments, we updated the parameters of $G$ and $D$ using different mini-batch data like the ways they do in image tasks. In other words, the parameters of $D$ were updated using one mini-batch data and then the parameters of $G$ were updated using a new mini-batch data. We found that this training strategy---LSTM+GAN+Res (DB) in the second-to-last row of Table~\ref{tab:res_gan}---was quite unstable in our experiments and we always achieved results worse than the non-adversarial training (e.g., LSTM+Res) as shown in Table~\ref{tab:res_gan}. Instead, when we tried to update the parameters of $G$ and $D$ using the same mini-batch data, we achieved consistently better results (LSTM+GAN+Res in Table~\ref{tab:res_gan}). We believe that this strategy is essential in making our GAN approach performing well. Adding residual connections works for most cases. LSTM+GAN+Res lowered the MCT AM CER from 15.35\% down to 13.14\% with 14.4\% relative CER reduction for the Real set and lowered the MCT AM CER from 14.03\% down to 11.40\% with 18.75\% relative CER reduction for the Simu set. Finally, we also find the performance of LSTM+GAN+Res+CD is worse than LSTM+GAN+Res. This means that adding the reverberant spectrogram as a condition to D is useless to the dereverberation performance.

\section{Summary}\label{section:conclusion}
In this paper, we provide a deep investigation of GAN in mapping-based speech dereverberation for robust speech recognition. In the selection of the generator network, we find that LSTM achieves superior performance, while adding residual connections (ResNets) in deep LSTMs can further boost the performance. In the use of GAN, we find that it is essential to update the generator and the discriminator using the same mini-batch data during model training; and using reverberant spectrogram as a condition to the discriminator may degrade the performance. With the above findings, we are able to achieve 14\%$\sim$22\% relative CER reduction in ASR as compared with a DNN baseline, while the SEGAN baseline even does not work on the ASR task. In the future, we plan to further explore the use of GAN in more adverse conditions (both reverberant and noisy) and try to combine the framework with joint-training strategy to further improve the ASR performance.

\section{Acknowledgements}\label{section:acknowledgment}
The authors would like to thank Shan Yang from Northwestern Polytechnical University, Dr. Bo Li from Google and Dr. Bo Wu from Xidian University for their helpful comments and suggestions on this work. The research work is supported by the National Key Research and Development Program of China (Grant No.2017YFB1002102) and the National Natural Science Foundation of China (Grant No.61571363).

%\clearpage
\bibliographystyle{IEEEtran}

\bibliography{interspeech2018-GAN}

\end{spacing}
\end{document}